\documentclass[letterpaper,twocolumn,letter]{jpsj3}
\usepackage{txfonts}
\usepackage{color}
\usepackage{braket}
\usepackage{bm}
\newcommand\ti\textit
\newcommand\tb\textbf
\newcommand\tx\text
\newcommand\lt\left
\newcommand\rt\right
\newcommand\ve\varepsilon
\newcommand\mr\mathrm
\newcommand\rng\rangle
\newcommand\lng\langle
\newcommand\vphi\varphi
\newcommand\ua\uparrow
\newcommand\da\downarrow
\newcommand\pd\partial
\newcommand\ov\overrightarrow
\newcommand\ol\overline
\newcommand\wt\widetilde
\newcommand\vv\vec

\title{Revisiting $^{9}$Be Nuclear Magnetic Resonance in UBe$_{13}$: Itinerant-Localized Duality and Possible Fermi Surface Reconstruction at High Magnetic Field }

\author{Rintaro Matsuki, Shoko Minami, Hisashi Kotegawa, Hisatomo Harima,\\
Yoshinori Haga$^2$,  Etsuji Yamamoto$^2$, Yoshichika \={O}nuki$^3$, and Hideki Tou\thanks{tou@crystal.kobe-u.ac.jp} }

\inst{Department of Physics, Graduate School of Science, Kobe University, Kobe 657-8501, Japan. \\
$^2$Advanced Science Research Center, Japan Atomic Energy Agency, Tokai 319-1195 Japan.\\
$^3$ Department of Physics, Graduate School of Science, Osaka University , Toyonaka, Osaka 560-0043, Japan.
}

\abst{ 
 We report on new results of $^{9}$Be nuclear magnetic resonance (NMR) measurements conducted on a single crystal of the heavy fermion superconductor UBe$_{13}$. Our previous 2007 study [J. Phys. Soc. Jpn. {\bf 76} 204705 (2007)] determined NMR and electric field gradient (EFG) parameters that successfully reproduced the NMR spectra at low magnetic fields. However, these parameters did not accurately describe the angular dependence of the NMR spectra at high magnetic fields. To address this discrepancy, we have now performed a more comprehensive investigation, measuring the magnetic field dependence of the $^{9}$Be-NMR spectra across a field range of 0.5 T to 8 T, as well as the magnetic field angle dependence at 0.5 T and 6 T. Through detailed simulations that take into account the non-symmorphic space group of UBe$_{13}$, we have determined a new set of parameters capable of reproducing the complex NMR line profiles observed at high magnetic fields. Notably, our analysis reveals the significant influence of classical dipolar fields.  A comparison between the Knight shift (KS) and the classical dipolar shift provides microscopic supporting evidence for the nature of an itinerant-localized duality in UBe$_{13}$. Furthermore, the magnetic field dependence of the KS exhibits anomalies around 6 T, suggesting a reconstruction of a part of the multiple Fermi surfaces in the high magnetic field region. 
}

\begin{document}
\maketitle
\section{Introduction}
UBe$_{13}$ is a well-known heavy fermion superconductor discovered by Ott {\it et al.} in 1983.\cite{Ott1} Extensive experimental and theoretical studies have suggested the realization of a spin-triplet superconducting state in this material.\cite{Remenyi,Brison,Ott2,Maple} However, the complexity inherent in its normal and superconducting state properties has thus far prevented a complete understanding of both states. To elucidate the nature of both the normal and superconducting states, detailed investigations from a microscopic perspective are crucial, in addition to bulk measurements. 

Initial $^9$Be nuclear magnetic resonance (NMR) studies were carried out  in the late 1980s, primarily on powdered samples and randomly oriented single crystals.\cite{MacLaughlin,Tien,Clark,Ahrens} These studies revealed anomalous enhancements in the $^9$Be NMR relaxation rate and the $^9$Be Knight shift (KS) attributed to the heavy Fermi liquid formation.  However, precise determination of NMR parameters, such as the KS tensor and the electric field gradient (EFG) tensor at the Be sites, was challenging due to the overlapped $^9$Be-NMR spectra, resulting  primarily from the low local symmetry at the Be sites forming an icosahedral cage. In 2007, we reported precise $^9$Be-NMR spectra and their analyses for a single-crystal of UBe$_{13}$. We evaluated NMR and EFG parameters, including quadrupole frequencies $\nu_Q$, asymmetry parameters $\eta$, hyperfine coupling constants $A$, and classical dipolar fields $H_{\rm dip}$, using the angular and temperature dependences of NMR spectra at $\mu_0H \approx 1$ T.\cite{Tou1}

However, these parameters failed to reproduce the angular dependence of the observed NMR line profiles at magnetic fields exceeding 3 T, despite providing a  rough fit to the data for $H \| [001]$.\cite{Morita}  This discrepancy was subsequently attributed to the oversight of the non-symmorphic space group of UBe$_{13}$, which features a $c$-glide symmetry. Notably, at high magnetic fields above 3 T, the influences of the EFG tensor and the classical dipolar field become crucial for accurately explaining the complex $^9$Be-NMR line profiles. In this paper, we present results from precise angle-resolved $^9$Be-NMR experiments. New NMR results and their reanalysis, which accounts for the crystal symmetry, are reported in {\S}3. From the field strength and field angle dependences of the KSs, we will discuss an itinerant-localized duality ({\S}4.1), a possible Fermi surface reconstruction  ({\S}4.2), and emergence of magnetic correlation around 6 T ({\S}4.3) in UBe$_{13}$.

\begin{figure}[htbp]
\begin{center}
\includegraphics[width=1\linewidth]{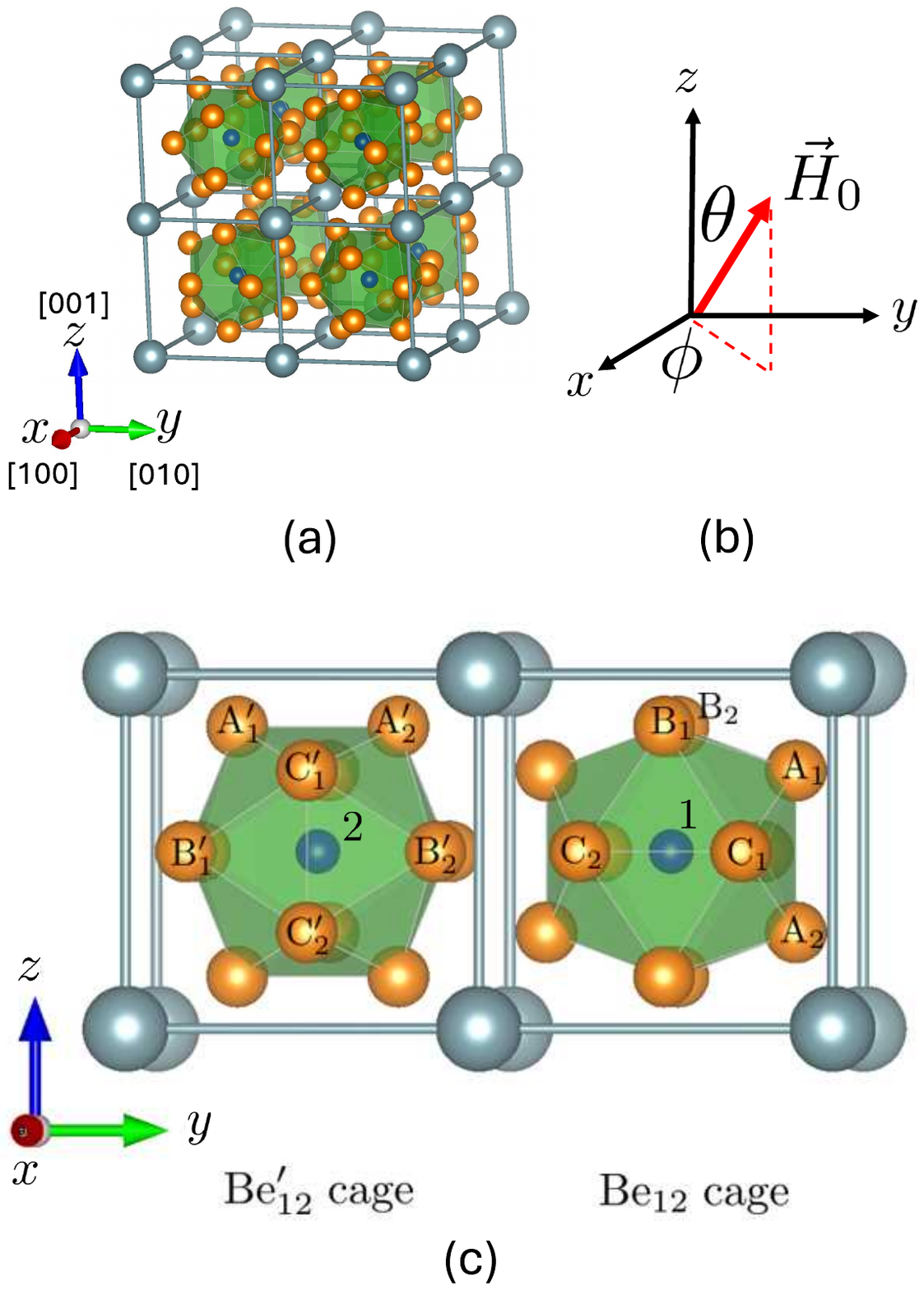}
\end{center}
\caption{(Color online) (a) Schematic view of the crystal structure of UBe$_{13}$.  (b) The direction of applied magnetic field in crystal axis coordinate frame $\Sigma^c$ using polar angles ($\theta, \phi$).  (c)  Arrangements of Be$_{12}$ and Be$_{12}^\prime$ cages. For example, legends, A$_1$ and  A$_1^\prime$  correspond to the Be$_{\rm II}$(A$_1$)  and  the Be$_{\rm II}$(A$_1^\prime$) sites, respectively.
}
\end{figure}

\section{Experimental Procedure}
Single crystals of UBe$_{13}$ with $T_c=0.85$ K were grown by the Al-flux method and characterized to be a single phase by Laue X-ray backscattering and powder X-ray diffraction.\cite{Haga} UBe$_{13}$ crystallizes in the cubic NaZn$_{13}$-type structure [``nonsymmorphic'' space group $Fm\bar{3}c$ (No. 226), $O_h^6$] (Fig. 1(a)). The structure contains two crystallographically inequivalent Be sites. The Be$_{\rm I}$ site has cubic symmetry ($m\bar{3}.$) at the $8b$ position: (0, 0, 0); (1/2, 1/2, 1/2). The Be$_{\rm II}$ site forms the vertices of an icosahedron and has lower symmetry ($m..$) at the $96i$ position: ($0, y, z$); etc., and thus experiences a finite EFG, $V_{\alpha\beta}=\frac{\partial^2V}{\partial \alpha \partial \beta}$, where $V$ is the electrostatic potential and $\alpha, \beta=x, y, z$. The single crystalline sample used for the $^{9}$Be NMR measurements was a cuboid with dimensions $L([100])\times W([010])\times H([001])\approx2\times2\times2$ mm$^3$. $^{9}$Be NMR measurements ($I=3/2$, $\gamma_{\rm n}/2\pi=5.9833$ MHz/T) were carried out by a conventional pulsed NMR spectrometer using  a 9 T solenoid superconducting magnet (JASTEC, 1 ppm/10mm DSV). The magnetic field is stable with a decay of less than 1 G per week in persistent supercurrent mode. Frequency-swept $^{9}$Be NMR spectra were obtained by Fourier transform (FT) of the spin-echo signal following a standard spin-echo pulse sequence, with a frequency step of 5 kHz at a fixed magnetic field. Field-angle dependent NMR measurements were conducted at 0.5, 1, and 6 T using a top-loading NMR probe equipped with a two-axis sample rotator. The magnetic field was calibrated using the $^{27}$Al resonance ($\gamma_n/2\pi=11.094$ MHz/T, $^{27}K=0.161$ \% at 4.2 K) from an aluminum reference sample.\cite{Metallic} 

\begin{figure}[htbp]
     \centering
     \includegraphics[width=1\linewidth]{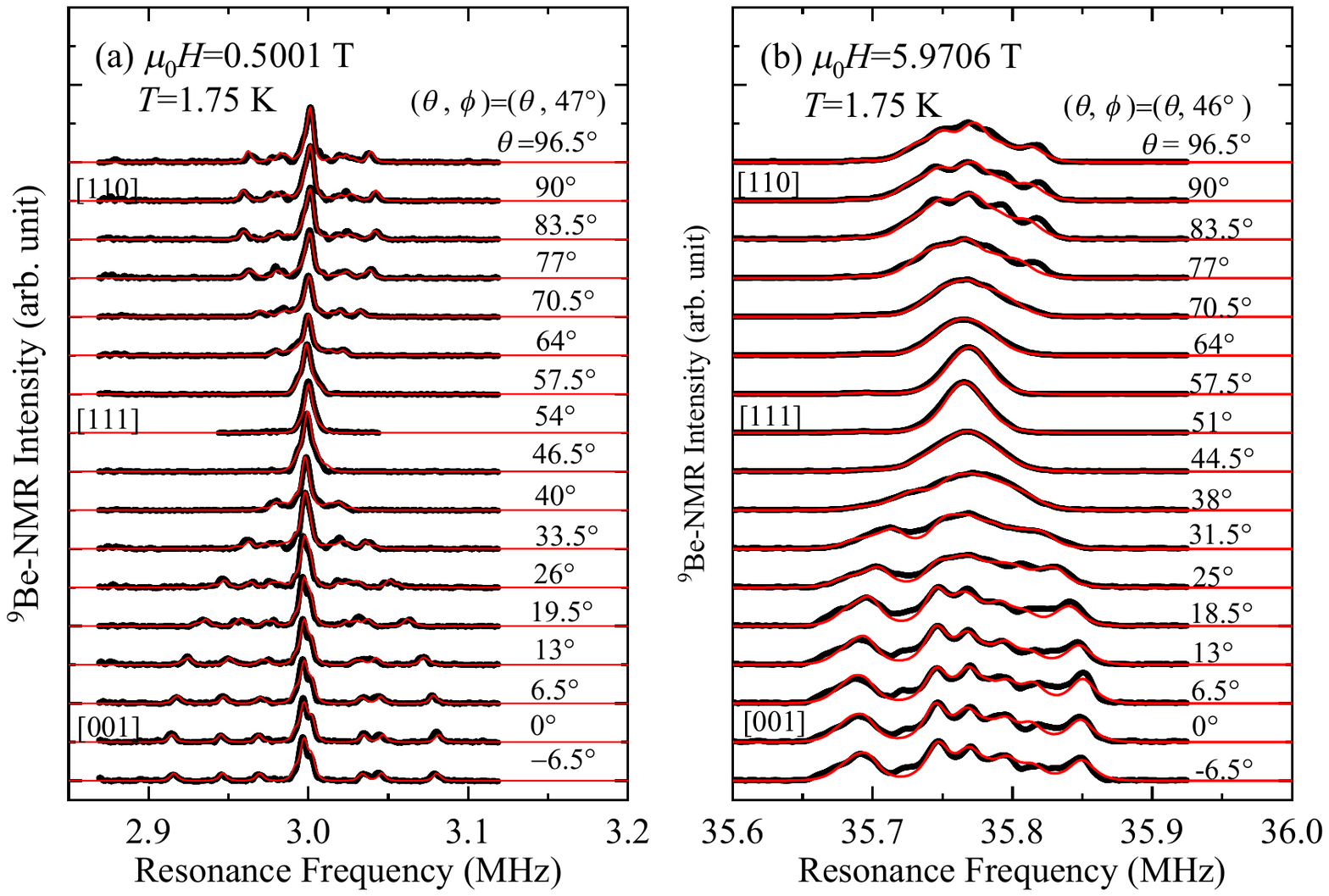}
     \caption{(Color online)
Magnetic field-angle dependence of the $^9$Be NMR spectra, measured at 1.75 K, for applied magnetic fields of  (a) $\mu_0H=0.5$ T and (b) 6 T.  The field angle was varied from the [001] to [110] crystallographic direction. The red solid lines are calculations with the nuclear spin Hamiltonian (see text).
 }
     \label{ir_spc_K}
\end{figure}

\section{Results}
\subsection{$^9$Be-NMR spectra under different magnetic field orientations}
Figures 2(a) and 2(b) present the field-angle dependence of the $^9$Be NMR spectra, measured at 1.75 K, for applied magnetic fields of  $\mu_0H\approx 0.5$ and 6 T, respectively. The field angle was varied from the [001] to [110] crystallographic directions as illustrated in Fig. 1(b).  At $\approx$0.5 T, the spectrum exhibits a remarkably narrow full width at half maximum (FWHM) of approximately 10 kHz, with no discernible additional signals, indicating the high crystalline quality and homogeneity of the single crystal at a microscopic level. The line profile at $\approx$ 0.5 T  is well reproduced using previously reported EFG and NMR parameters,\cite{Tou1} which considered only a single Be$_{12}$ cage. These parameters account for the resonance signal from the Be$_{\rm I}$ site, which experiences a cubic local symmetry and thus zero EFG, and the resonances from three crystallographically inequivalent Be$_{\rm II}$ sites located at the vertices of the icosahedral cage. These Be$_{\rm II}$ sites are denoted as Be$_{\rm II}$(A), Be$_{\rm II}$(B), and Be$_{\rm II}$(C), each characterized by a finite EFG tensor.\cite{Tou1}  For an arbitrary field direction, a fourfold Be$_{\rm II}$(X) site (X=A, B, and C) splits into two inequivalent twofold Be$_{\rm II}$(X$_1$) and Be$_{\rm II}$(X$_2$) sites due to the rotation of the maximum principal component of the  EFG tensor within the mirror plane (e.g.,  $yz$-plane for Be$_{\rm II}$(A) site). In contrast, a significant deviation is observed between these parameters and the data obtained at $\approx$6 T.  This discrepancy arises from the neglect of the non-symmorphic space group  ($Fm\bar{3}c$) of UBe$_{13}$ in the previous simulation. The presence of the $c$-glide symmetry within the $Fm\bar{3}c$, operating across diagonal planes such as $ (1\bar{1}0)$, necessitates the inclusion of contributions from the adjacent Be$_{12}^\prime$ cage, as illustrated in Fig. 1(c). This $c$-glide operation results in a 90-degree relative rotation of the icosahedral phase between neighboring Be$_{12}$ and Be$_{12}^\prime$ cages. Consequently, the EFG and KS tensors of the Be$_{12}^\prime$ cage are related to, but distinct from, those of the Be$_{12}$ cage by this rotational transformation. Hereafter, the Be$_{\rm II}$ sites in the Be$_{12}^\prime$ cage are denoted as Be$_{\rm II}$(X$^\prime_i$) (X=A, B, and C; $i=1, 2$),  as depicted by legends (e.g., A$^\prime_1$) in the figure.

\subsection{$^9$Be-NMR spectra under different magnetic field strengths}
Figures 3(a) and 3(b) show the magnetic field dependences of the $^{9}$Be NMR spectrum for $H \| [001]$ and  $H\| [111]$, respectively.   With increasing magnetic field, the NMR line profile changes significantly, and the linewidth broadens to $\approx$40 kHz. This broadening is attributed to the combined effects of  KS and  EFG tensors, and  the classical dipole field. However, the line profile cannot be reproduced by the previously obtained parameters, where  the classical dipolar field was calculated and provided only for $\mu_0H \approx 1$ T, assuming the uranium moment assumed to be 0.021 $\mu_B$/U.\cite{Tou1}  The magnetization exhibits an approximately linear response to the magnetic field up to high magnetic fields exceeding 20 T\cite{Detwiler}. Therefore, in our current magnetic field dependence measurements,  we must precisely evaluate the influence of the classical dipolar field by considering the changes induced by the field-dependent magnetization on the NMR spectrum.

\begin{figure}[htbp]
     \centering
     \includegraphics[width=1\linewidth]{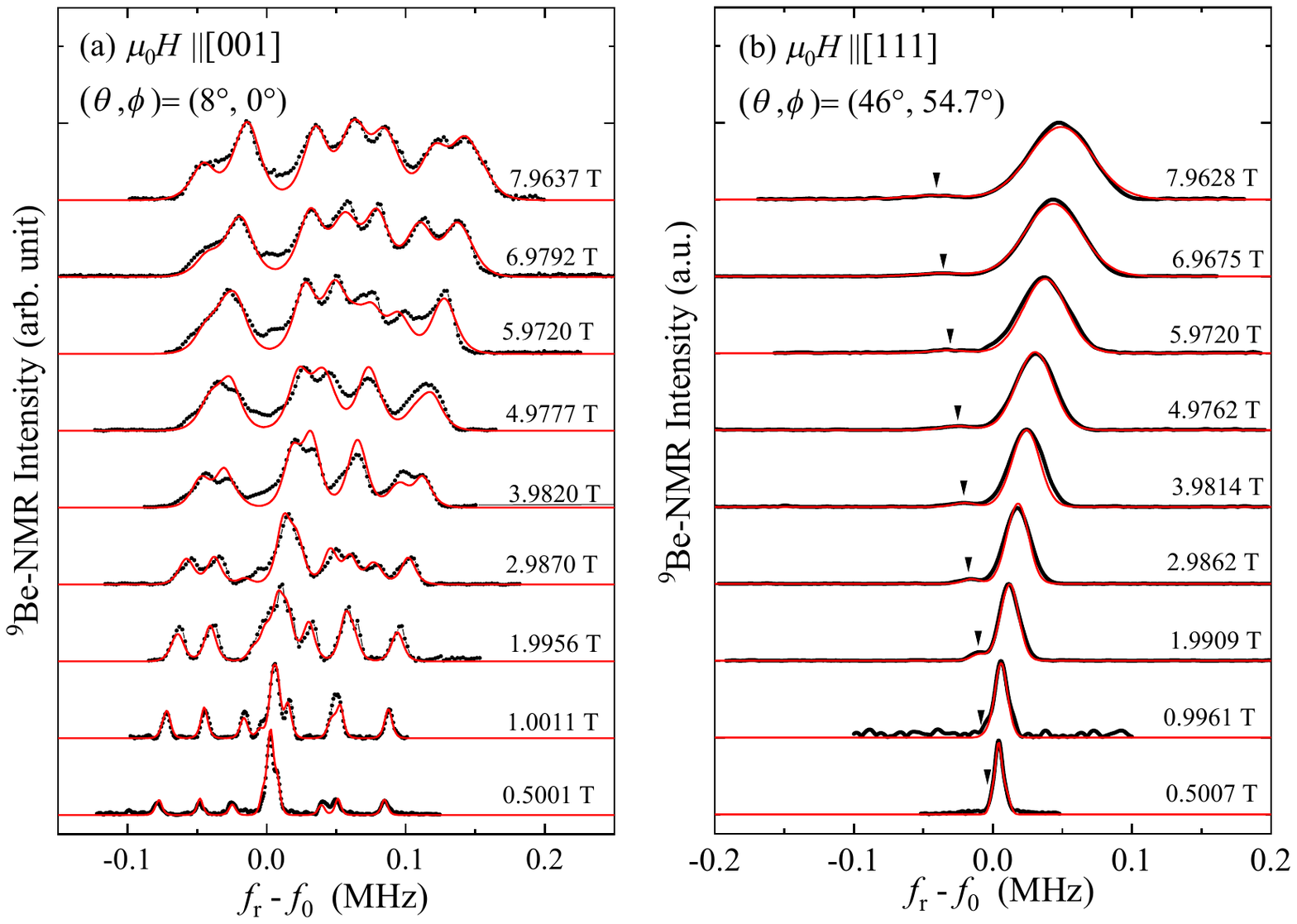}
     \caption{(Color online)
Magnetic field dependence of $^9$Be NMR spectra for (a) $H\|$[001] and (b)  $H\|$[111]. Triangle symbols indicate the resonance frequency of the Be$_{\rm I}$ site. The red solid lines are calculations with the nuclear spin Hamiltonian (see text).
 }
     \label{ir_spc_K}
\end{figure}

\subsection{$^9$Be-nuclear spin Hamiltonian}
 In order to reproduce the complex NMR line profiles at high magnetic fields, simulation calculations were performed by precisely diagonalizing the 4$\times$4 nuclear spin Hamiltonian matrices for $^{9}$Be at 14 different sites of the Be$_{12}$ and Be$_{12}^\prime$ cages. Here, considering the rotational symmetry with respect to the magnetic field, we take account of a total of 12 Be$_{\rm II}$ sites, which are the six sites in the Be$_{12}$ cage and the six sites in the Be$_{12}^\prime$ cage, along with two sites for Be$_{\rm I}$.

Firstly, we define nuclear spin Hamiltonian for the Be$_{\rm II}$(A$_1$) site at the position (0.25,0.4261,0.3641) in the crystal coordinate frame, $\Sigma^{\rm c} $ defined by the crystal axes $(x,y,z)$,  as shown in Fig. 1(b). Then, the Hamiltonian  is written as {
\begin{align}
{\cal H}_{\rm II}({\rm A_1})=&-\gamma_{\rm n} \hbar\vec{I}\cdot\lt(\tilde{E}-K_{\rm II}^{\rm iso}\tilde{E}+\tilde{K}_{\rm II}^{\rm an}+\tilde{D}\rt)\cdot\vec{H}  \notag \\ 
+&\frac{eQ}{6I(2I-1)}\sum_{\alpha,\beta=x,y,z}V_{\alpha\beta}\lt[\frac{3}{2}(I_\alpha\cdot I_\beta)-\delta_{\alpha\beta}I^2\rt].
\end{align}
}

The first line of Equation (1) represents the magnetic interactions, and the second line represents the electric interactions. The first term of the magnetic interactions is the Zeeman interaction between the Be nuclear spin and the external magnetic field, $\vec{H}$. Here, $\tilde{E}$ is the second-rank unit tensor, and  the direction of the external magnetic field is expressed by $\vec{H}=(H_{x}, H_{y}, H_{z})=H(\cos\theta\sin\phi, \sin\theta\sin\phi, \cos\theta)$, using the polar angle $\theta$ and the azimuthal angle $\phi$ as shown in Fig. 1(b).  The second  term, $K_{\rm II}^{\rm iso}$,  is the isotropic part of the KS resulting from the sum of core polarization and the Fermi contact hyperfine interactions. The third term, $\tilde{K}_{\rm II}^{\rm an}$, is the anisotropic part of the KS originating from the spin dipolar interactions, which are attributed to the local-spin-density at the Be $2p$ orbital. In the present case, for Be atoms, the $2s$ and $2p$ states at the Fermi level in the conduction band hybridize with the $5f/6d$ states of U ions, carrying spin density that gives rise to the transferred hyperfine field on Be nuclei. \cite{Tou1} $\tilde{K}_{\rm II}^{\rm an}$ is written as   
\begin{equation}
\tilde{K}_{\rm II}^{\rm an}=
\begin{bmatrix}
K_{xx}  & 0 & 0  \\
0 & K_{yy}  & K_{yz}  \\
0 & K_{yz}  & K_{zz}
\end{bmatrix}
,
\end{equation} 
according to the local symmetry at the Be$_{\rm II}$(A$_1$) site.  The fourth term of the first line in Eq. (1) is due to the classical point dipolar field and this contribution can be written as following  tensor form, 
\begin{equation}
\tilde{D}=\frac{M}{H}\sum_j\frac{\mu_B}{r_j^5}
\begin{bmatrix}
3x_j^2-r_j^2  & 3x_jy_j & 3x_jz_j  \\
3y_jx_j & 3y_j^2-r_j^2  & 3y_jz_j  \\
3z_jx_j & 3z_jy_j  & 3z_j^2-r_j^2 
\end{bmatrix}
,
\end{equation}
where $M$ is the magnetization and $\vec{r}_j =(x_j,y_j,z_j)$ is the position vector of the uranium ions with respect to the Be$_{\rm II}$(A$_1$) site.
Note that the Van Vleck contribution to the KS is negligible according to the previous report\cite{Tou1}. 
  
  The second line of Equation (1) is  the nuclear quadrupole interaction. $Q$ is the nuclear quadrupole moment, and $V_{\alpha\beta}$ is the component of the EFG tensor $\tilde{V}_{\alpha\beta}$ as mentioned in the previous section.  Here  $\tilde{V}_{\alpha\beta}$ is written as   
\begin{equation}
\tilde{V}_{\alpha\beta}=
\begin{bmatrix}
V_{xx}  & 0 & 0  \\
0 & V_{yy}  &V_{yz}  \\
0 & V_{yz}  & V_{zz}
\end{bmatrix}
,
\end{equation}
according to the local symmetry at the Be$_{\rm II}$(A$_1$) site. The KS and EFG tensors for the other 11 Be$_{\rm II}$ sites can be obtained by a $zxz$-active rotational transformation, $\tilde{T}^\prime=O\tilde{T}O^T$, of the  KS and  the EFG tensors ($\tilde{T}=\tilde{K}_{\rm an}, \tilde{V}_{\alpha\beta}$) of the Be$_{\rm II}$(A$_1$) site. Here $O$ and $O^T$ are the $zxz$-active rotation matrix and its transpose, respectively. Corresponding dipolar tensors are calculated for each position.

As for the Be$_{\rm I}$ site, due to the local symmetry of $m\bar{3}.$ (cubic), it is not necessary to consider anisotropic KSs, EFG, or classical dipolar magnetic fields. {Then, the nuclear spin Hamiltonian is simply written as: 
\begin{align}
{\cal H}_{\rm I}&=-\gamma_{\rm n}  \hbar(1+K_{\rm I}^{\rm iso}) \vec{I}\cdot \vec{H}.
\end{align}
}
Thus, simulation calculations were performed for the Hamiltonian considering 12 Be$_{\rm II}$ and 2 Be$_{\rm I}$ sites written as, 

\begin{align}
{\cal H}_0&=\sum_{i}{\cal H}_{\rm II}(i)+\sum_{j}{\cal H}_{\rm I}(j), \\ 
i=&{\rm A_1, A_2, B_1,} \cdots, {\rm B_2^\prime, C_1^\prime, C_2^\prime}; \mbox{\hspace{3mm}} j=1, 2.  \notag
\end{align}
Here, the transformation rule between Be$_{\rm II}$ tensors, $\tilde{T} (=\tilde{K}, \tilde{V}_{\alpha\beta}$), in the crystal axis coordinate frame, $\Sigma^{\rm c} $, and the tensors $\tilde{T}^{\rm P}$ in the principal axis coordinate frame, $\Sigma^{\rm P}$, of principal axes ($X, Y, Z$) is given by  $\tilde{T}^P=v^T\tilde{T}v$.  where $v$  is a matrix whose column vectors are eigenvectors, and the subscript ``P'' refers to the principal axis systems of the Zeeman interaction (P=Z), the hyperfine interaction (P=hf), the dipolar interaction (P=D), and the nuclear quadrupole interaction (P=EFG).\cite{Cohen} The nuclear spin  $\vec{I}$ and magnetic field $\vec{H}$  at the Be$_{\rm II}$(A$_1$) site are also  transformed to the principal frame $\Sigma^{\rm P}$ by $\vec{I}^{\rm P}=R\vec{I}$ and $\vec{H}^{\rm P}=R\vec{H}$, respectively, where $R=v^T$ is a rotation matrix.

\subsection{$^9$Be-NMR spectral analysis below 2 T}
In general, simulating a spectrum using the Hamiltonian (Eq. (6)) involves numerous parameters. However, the number of parameters can be effectively reduced by utilizing the simulations of the spectra measured  at 0.5 T, as follows: The previously determined $\nu_Q$ and $\eta$ can be used as reliable initial values for the subsequent analysis. Indeed, the maximum splitting of the satellite lines occurs when the magnetic field is aligned parallel to the principal axis of the EFG tensor ($V_{ZZ}$ and $Z$ is the direction of the maximum principal axis of the $\Sigma^{\rm EFG}$ frame), strictly yielding a splitting of $2\nu_Q$.  Consequently, $\nu_Q$ is precisely determined to be 83.7$\pm$0.05 kHz, which is consistent with the previous result within the experimental error.

Conversely, when the magnetic field  is applied strictly perpendicular to the $V_{ZZ}$, the splitting width of the satellite lines exhibits an angular dependence within the plane perpendicular to the $V_{ZZ}$. The splitting is  described by $\nu_Q\eta(1+\cos2\alpha)$ within the framework of the first-order perturbation theory,\cite{Metallic} where $\alpha$ represents the azimuthal angle of the magnetic field direction with respect to the $Z$ axis of the $\Sigma^{\rm EFG}$ frame. This angular dependence indicates that the EFG tensor deviates from axial symmetry and $\eta$ can be estimated to be $\approx$0.2, confirming the validity of the previous estimation.
Ultimately, the refined values for $\nu_Q$ and $\eta$ are determined to be  83.7$\pm$ 0.05 kHz and 0.21$\pm$0.005, respectively. Correspondingly, the principal components of the EFG tensor are found to be $(V_{XX},V_{YY},V_{ZZ})=(-0.51589,-0.79286, 1.3088)\times10^{20}$ V/m$^2$, where we define $|V_{ZZ}|\geq|V_{YY}|\geq|V_{XX}|$. These experimentally determined values are approximately 70 $\%$ of those predicted by band structure calculations.\cite{Tou1}

Furthermore, for $H \| [110]$, the rotation angle of the principal axes of the EFG tensor within the mirror plane (e.g., the $yz$-plane for the Be$_{\rm II}$(A) site) can be estimated. If the principal axes of the EFG tensor were aligned precisely with specific crystallographic axes, the spectrum would exhibit three distinct pairs of satellite lines symmetrically positioned around the central frequency. However, if the principal axes are rotated by an angle within the mirror plane, the resonance lines from the non-equivalent Be$_{\rm II}$(B) and Be$_{\rm II}$(C) sites can merge. As clearly shown in Fig. 2(a), the obtained line profile for $H \| [110]$ demonstrates that $V_{ZZ}$ and $V_{YY}$ are rotated away from the crystal axes within the $yz$-plane, but this rotation angle is limited to approximately $2.5\pm0.5$ degrees and is therefore quite small. Note that the obtained value is consistent with the previous band structure calculation.\cite{Tou1}

The isotropic KS, $K^{\rm iso}$ can be determined independently from the simulation of each spectrum, regardless of the spectral broadening or shape.  This is because multiple Be$_{\rm II}$ sites connected through the rotational transformations can be observed in just one measurement. The isotropic KSs for Be$_{\rm II}$ and Be$_{\rm I}$ at $\approx 0.5$ T are obtained to be $K_{\rm II}^{\rm iso}=0.095\pm0.008$ \% and $K_{\rm I}^{\rm iso}=-0.08\pm0.01$ \%, respectively, which are almost the same as the previous values.

Finally, anisotropic KS tensors and classical dipolar tensors are determined from the magnetic field dependence of the spectra shown in Fig. 3. Here, the principal values of the classical dipolar tensors are numerically calculated using Eq. (3) with the field dependence of the magnetization, $M(H)\approx 0.017\times H$  $\mu_B/$U. The summation was taken over the uranium moments at 4913 uranium sites. Consequently, the dipolar field is obtained to be $(H_{XX}^{\rm D},H_{YY}^{\rm D},H_{ZZ}^{\rm D})=(-1.70, -10.1 ,11.8)$ G at  1 T. This value corresponds to the classical dipolar tensor, $\tilde{D}$, to be $(D_{XX},D_{YY},D_{ZZ})=(-0.017,-0.101,0.118)$ \% with respect to the frame of the dipole field, $\Sigma^{\rm D}$. In this calculation, the classical dipolar field was calculated by the localized moment of the U site, where the magnitude of the moment was taken as $\approx10$ \% smaller than the experimental value.\cite{Tou1} 

\begin{figure}[htbp]
     \centering
     \includegraphics[width=1\linewidth]{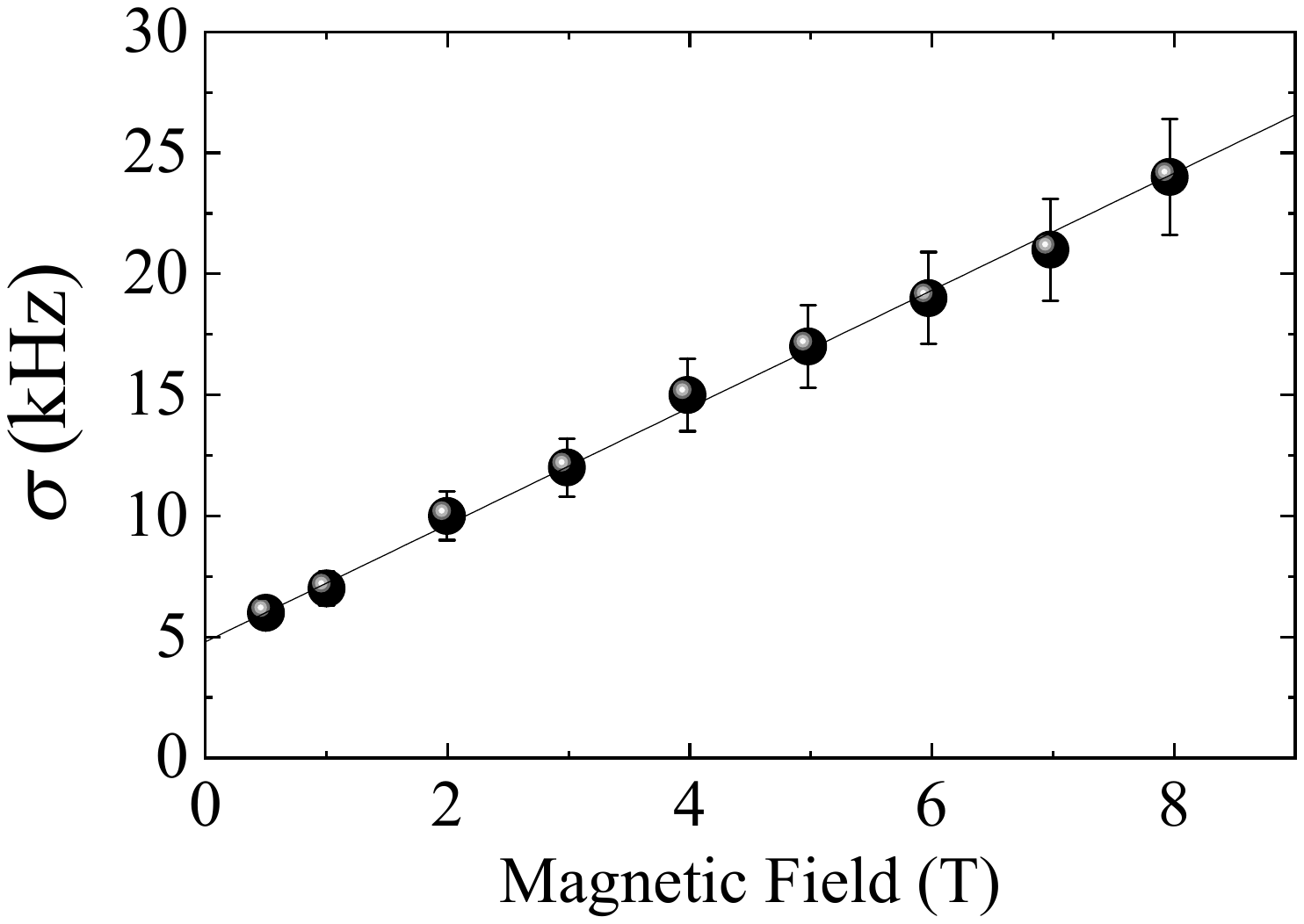}
     \caption{(Color online)
Magnetic field dependence of the linewidth of the convolution function for Be$_{\rm I}$ and Be$_{\rm II}$ NMR lines.
 }
     \label{ir_spc_K}
\end{figure}

Figure 4 illustrates the magnetic field dependence of the linewidth $\sigma$ of the convolution function used in the spectral analysis. The NMR spectral linewidths  for both $H \| [001]$ and $H \| [111]$  exhibit a linear dependence on the magnetic field within the measured field range. This line broadening  with increasing $H$ is attributed to the anisotropy of the dipolar field.  To reproduce the magnetic field and angular dependences of the line profile, the principal values of the anisotropic KS tensor measured at 0.5 T are corrected to $(K_{XX},K_{YY},K_{ZZ})=(0.006,0.194,-0.200)$ \% with respect to the frame of the Knight shift, $\Sigma^{\rm hf}$, where we define $|K_{ZZ}|\geq|K_{YY}|\geq|X_{XX}|$. The resulting KS parameters obtained at 0.5 T are listed in second column in Table I. Below 2 T, $K_{\rm II}^{\rm iso}$ and $K_{\rm I}^{\rm iso}$  do not depend on the magnetic field strength or magnetic field angle.  The results of simulations thus obtained are shown by the red solid lines in Figs. 2 and 3.

\subsection{$^9$Be-NMR spectral analysis above 2T}

{
Above 2 T, both $K_{\rm II}^{\rm iso}$ and $K_{\rm I}^{\rm iso}$  become dependent on the magnetic field strength and its angle, and cannot be reproduced using the parameters derived in the low-field region. Therefore, we added components for the magnetic-field-induced anisotropic KS  ($\tilde{K}_{\rm II}^{\rm ind}$ for Be II and $\tilde{K}_{\rm I}^{\rm ind}$ for Be I) to the analysis, modifying Equation (6) as follows:
\begin{align}
{\cal H}&={\cal H}_0-\gamma_{\rm n}  \hbar \vec{I}\cdot\lt(\tilde{K}_{\rm II}^{\rm ind}+\tilde{K}^{\rm ind}_{\rm I}\rt)\cdot \vec{H}.
\end{align}
Here, the Hamiltonian, ${\cal H}_0$, is reproducible with parameters obtained from magnetic fields below 2 T. We assumed that the magnitude of the anisotropic KS components of  $\tilde{K}_{\rm II}^{\rm an}$  is independent of the magnetic field angle. This is because trial-and-error simulations performed at various magnetic field angles showed a variation of less than 10 \% from the values obtained below 2T, and this change is comparable to the overall experimental errors. Furthermore, when $H\|[111]$, the magnetic field is applied at an angle close to the magic angle with respect to the principal axis of the anisotropic KS at all sites. This makes it impossible to accurately determine the anisotropic KS value. In fact, as shown in Table I, the simulation results revealed that $K_{\rm II}^{\rm an}$  also depends on the magnetic field, but this change is only about a 5 \% increase, which is on the same order as the experimental error. Therefore, we considered the magnetic field dependence of the anisotropic shift to be negligible.

The magnetic field dependences of the obtained field-induced components, $K_{\rm II}^{\rm ind}$ and $-K_{\rm I}^{\rm ind}$, are shown in Figs. 5(a) and 5(b) for magnetic fields applied along the cubic principal axes, [001] and [111]. Furthermore, the anisotropy of the field-induced component at 6 T becomes pronounced  between $H\|$[001] and $H\|$[111]. Figures 6(a) and 6(b) show the behavior of $K_{\rm II}^{\rm ind}$ and $-K_{\rm I}^{\rm ind}$ as the angle is changed from the [001] to [100] direction, with the magnetic field fixed at 6 T. For $H\|$[001] direction, the field-induced components are $K_{\rm II}^{\rm ind}\approx 0.04$ \% and $-K_{\rm I}^{\rm ind}\approx 0.07$ \%. In contrast, when $H\|$[111] direction, the components are $K_{\rm II}^{\rm ind}\approx 0.005$ \% and $-K_{\rm I}^{\rm ind}\approx 0.02$ \%, which are notably smaller than those for $H\|[001]$.

\begin{table}[htbp]
\caption{Obtained values of isotropic Knight shifts ($K^{\rm iso}$), anisotropic Knight shifts ($K_{\alpha\beta}$, $\alpha,\beta=X,Y,Z$), nuclear quadrupole frequency ($\nu_{\rm Q}$), asymmetry parameter ($\eta$), and magnetization ($\mu_{\rm U}$) calcuated  at around 0.5 T and 6 T, together with the previous values obtained at 1 T by the field-swept NMR.\cite{Tou1}}
\begin{tabular}{cccc} 
parameters& 0.5 T  & 6T & previous   \\ \hline\hline
     $ K^{\rm iso}_{\rm I}$  (\%)& $-0.08\pm 0.01$&$-0.08\pm 0.01$ & $-0.081\pm 0.02$\\ \hline
     $ K^{\rm iso}_{\rm II}$ (\%)& $0.095\pm0.008$ &$0.095\pm 0.008$ & $0.13\pm0.02$  \\ \hline
      $K_{XX} $(\%)                & $0.006\pm 0.008 $&$0.006\pm 0.008$& $0.01\pm 0.02 $\\
      $K_{YY} $(\%)                &$0.194\pm 0.008$  &$0.203\pm 0.008$&$0.06\pm 0.02$ \\
      $K_{ZZ} $(\%)                &$-0.200\pm 0.008$&$-0.205\pm 0.008$&$-0.07\pm 0.02$\\\hline
      $\nu_{Q}$  (kHz)            &$ 83.7\pm 0.05$ &$ 83.7\pm 0.05$&$ 84.0\pm 0.5$ \\
      $\eta$ &$ 0.21\pm0.005$ &$ 0.21\pm0.005$& $0.21\pm0.01$ \\ \hline
      $\mu_U  (\mu_B/{\rm U})\times H$&$0.0085$ &0.102 &0.021
\end{tabular}
\end{table}

\section{Discussions}
\subsection{Itinerant-localized duality}

As shown in Table I,  the isotropic KSs for the Be$_{\rm I}$ and Be$_{\rm II}$ sites are $K_{\rm II}^{\rm iso}\approx0.095$ \% and $K_{\rm I}^{\rm iso}\approx-0.08$ \%, which are consistent with prior data for isotropic values. The anisotropic KSs ($K_{\rm II}^{\rm an}$) are three times larger than the previously obtained values.\cite{Tou1}  The order of magnitude of KSs ($\approx$0.1 \%) contrasts sharply with the negligible KS of metallic Be, which exhibits only a very small isotropic shift of $K^{\rm iso}\approx-0.003 $ \%.\cite{Barnaal} The KS enhancements are approximately 100 times larger compared to non-interacting Be metal, evidencing a heavy fermion state characterized by a large Fermi level density of states, $N^\ast(\ve_F)$, due to $c-f$ hybridization.\cite{Tou1} In particular, the site symmetry point groups for U site ($8a$) and Be site ($96i$) are 432 and 1, respectively, and both lack local spatial inversion symmetry. This can lead to a substantial parity hybridization between orbitals with different parity. In this context, significant transferred hyperfine interactions arising from mixing between Be $2s/2p$ orbitals and U $5f/6d$ orbitals enhance both {$K^{\rm iso}$ and $K^{\rm an}$.}

In this case, quasiparticle spin susceptibility $\chi^{\rm qp}$ and electronic specific heat coefficient $\gamma_{\rm el}$ are enhanced by the magnetic enhancement factor, $\tilde{\chi}_0$,  and the mass enhancement factor, $\tilde{\gamma}$, respectively, related by the Wilson Ratio $R\equiv\tilde{\chi}_0/\tilde{\gamma}$. The relationship between $\chi^{\rm qp}$ and $\gamma_{\rm el}$ in units per mole  is written as,\cite{Tou2}
\begin{eqnarray}
\chi^{qp}=\frac{2}{3}N_{\rm A}g_J^2J_{\rm eff}^2\mu_{\rm B}^2N^{\ast}(\varepsilon_{\rm F})\frac{\tilde{\chi}_0}{\tilde{\gamma}} =\frac{\gamma_{\rm el} g_J^2J_{\rm eff}^2\mu_{\rm B}^2}{\pi^2k_{\rm B}^2}R,
\end{eqnarray}
where $N_{\rm A}$ is the Avogadro's number, $g_J$ the Lande factor,  $J_{\rm eff}$ the effective spin,  $\mu_{\rm B}$ the Bohr magneton, $N^{\ast}(\varepsilon_{\rm F})$ the one-spin effective DOS at the Fermi level,  $k_{\rm B}$ the Boltzmann constant. The KS is calculated  to be $K^{\rm qp}=0.107$ \%, which agrees well with the observed value {$K^{\rm iso}_{\rm II}= 0.095\pm 0.008$\%.} Here we used  the relation of $K^{\rm qp}=\frac{A_{\rm iso}}{N_{\rm A}\mu_{\rm B}}\chi^{\rm qp}$, assuming  $g_J\sqrt{J(J+1)}=1.732$ and $R=1$, with parameters of the isotropic hyperfine coupling constant $A_{\rm iso}=436$ Oe/$\mu_{\rm B}$,\cite{Tou1} and the electronic specific heat coefficient $\gamma_{\rm el}=1000$ mJ/mol K$^2$\cite{Shimizu2}.

On the other hand, as mentioned in the previous section, the classical dipolar shift  tensor  of $(D_{XX},D_{YY},D_{ZZ})=(-0.017,-0.101,0.118)$ \%  is of a  comparable order to the KS tensors. If the U $5f$ electrons were itinerant, the magnetic moment should be reduced due to this itinerancy, and the influence of the dipolar magnetic field should be small.  The consistent interpretation of NMR spectra using both the KS and the classical dipolar field suggests that localized U moments persist even in the heavy fermion state.

 Theoretically, for an $f^1$ heavy electron state at $T\ll T_{\rm K}$,  where $T_{\rm K}$ is the Kondo temperature, the one-particle DOS spectrum of $f$-electrons is known to feature a Kondo resonance peak from itinerant quasiparticles near the Fermi level and two incoherent localized peaks  far from it due to the strong Coulomb interactions.\cite{Sakai}  While the U system is considered an $f^2$ configuration, making it more complex than the  $f^1$ case,  the simultaneous validity of the classical dipolar field and the heavy Fermi liquid relation serves as an experimental signature of  the ``{\it Itinerant-localized duality}'' in heavy electron systems.\cite{Kuramoto,Kuramoto2}

\subsection{Possible Fermi surface reconstruction around $H_K$}

Here we examine the magnetic field-induced component of the KS above 2 T. {Both $K^{\rm iso}_{\rm II}$ and $K_{\rm II}^{\rm an}$ remain constant within experimental error up to $\approx$ 2 T.} However, above this field, {the field-induced component of the KS, $K_{\rm II}^{\rm ind}$, appears  and exhibits an anisotropy between the [001] and [111] directions.}  As shown by red open circles in Fig. 5, for $H \| [111]$, the {both $K_{\rm II}^{\rm ind}$ and $-K_{\rm I}^{\rm ind}$}   show a modest increase up to $\approx$0.02 \% with increasing $H$.  In contrast, for $H \| [001]$  (blue closed circles), the respective values for both Be$_{\rm II}$ and Be$_{\rm I}$ sites increase significantly,  reaching maxima of 0.04 \% and 0.065 \% around 6 T,  followed by  decreases in the higher field region above 6 T.

Around 6 T, field-induced anomalies have been reported in Hall resistivity,\cite{Alekseevskii} magnetic torque,\cite{Schmiedeshoff} and thermoelectric power measurements,\cite{Shimizu3} despite the absence of discernible anomalies in the magnetic field dependence of specific heat or magnetization.   These anomalies are observed at different magnetic field strengths ranging between 3.5 T and 8 T, depending on the experimental probe.  The Hall effect measurements show a sign change of the Hall coefficient around 7 T, which has been phenomenologically described by a two-carrier (two-band) model.\cite{Alekseevskii}  The magnetic torque measurements revealed anomalies in the field range of 3 $-$ 5 T, interpreted in the context of a potential field-induced magnetic phase transition \cite{Schmiedeshoff}.   Conversely, the thermoelectric power data suggest a change in the conduction electron lifetime, $\tau_c$, around 7 T, possibly indicative of Fermi surface reconstruction at high fields \cite{Shimizu3}. Band structure calculations indeed predict small, closed hole Fermi pockets originating from $f$-electrons at the X point in UBe$_{13}$ \cite{Takegahara,Maehira}, and it has been speculated that these heavy Fermi surfaces may undergo modifications or vanish in the field range of 6-8 T.\cite{Shimizu3}

What is the origin of the anomalous magnetic response? {Generally, the KS in conventional metals should be independent of both the magnetic field orientation and magnitude. Therefore, the appearance of the anisotropic field-induced components of the KSs above 2 T, that exhibits anomalous enhancement around 6 T for $H \| [001]$,  deviates from the simple Fermi liquid picture.} 
As previously noted, the NMR linewidth, $\sigma$, for both $H \| [001]$ and $H\| [111]$ directions exhibits an almost linear-field dependence within experimental accuracy, lacking any singular behavior. This suggests that the anomalous field dependence of the KSs is not attributable to a magnetic phase transition. Furthermore, a simple two-band model fails to account for the magnetic field orientation-dependent $K^{\rm ind}$ around 6 T. 

 \begin{figure}[htbp]
     \centering
     \includegraphics[width=1\linewidth]{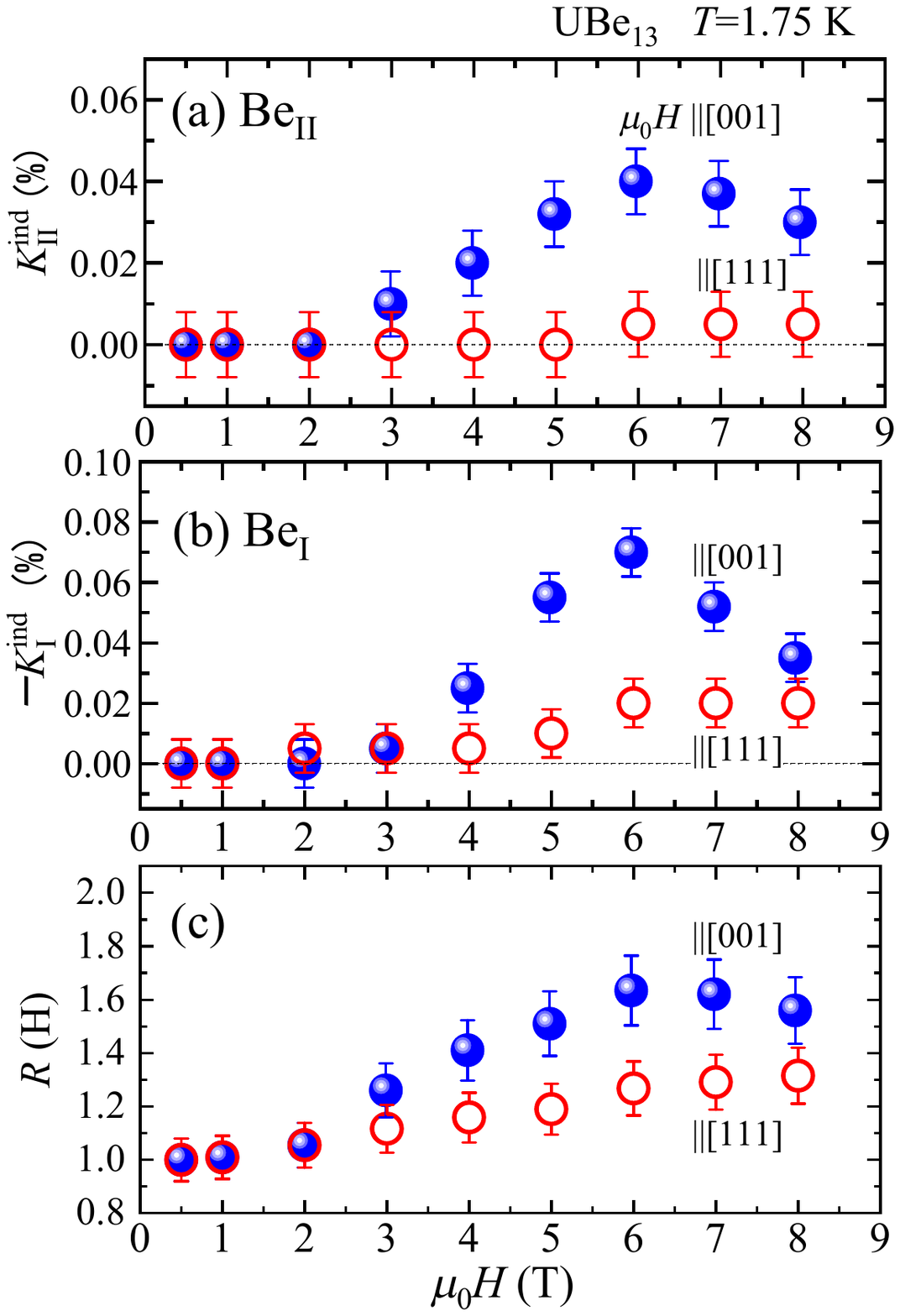}
     \caption{(Color online)
Magnetic field dependences of the field-induced components the KS,  (a) $K_{\rm II}^{\rm ind}$  for  Be$_{\rm II}$ site and (b) $K_{\rm I}^{\rm ind}$ for  Be$_{\rm I}$ site. Blue closed and red open circles are data for $H \| [001]$ and $H\| [111]$, respectively. Here, $K_{\rm II}^{\rm iso}=0.095$ \% and $K_{\rm I}^{\rm iso}=-0.08$ \%. (c) Magnetic field dependence of the Wilson ratio $R$ calculated using Eq.(9).
 }
     \label{ir_spc_K}
\end{figure}

Specific heat measurements reveal a monotonic and gradual decrease in the electronic specific heat coefficient, $C(H)/T$ (hereafter referred to as $\gamma_{\rm el}(H))$, with increasing magnetic field for both $H \| [001]$ and $H\| [111]$. This behavior is interpreted by the field suppression of the Kondo resonance.\cite{Shimizu2} However, above 2 T, anisotropy emerges; $\gamma_{\rm el}(H)$ decreases more rapidly with field for  $H \|[111]$ than  for $H \| [001]$, where the decrease remains gradual. 
Considering the specific heat data, the anomalous magnetic responses observed in Hall resistivity, magnetic torque, thermoelectric power, and the present KSs are likely related to the development of magnetic correlations driven by spin fluctuations around 6 T. { Within the framework of the heavy Fermi liquid regime, these correlations can be quantified by the magnetic enhancement factor, reflected in the Wilson ratio $R$. Since the magnetic field-induced KS and specific heat depend on the magnetic field strength and angle, we assume that $R$ depends on $H$ and $\theta$. Equation (8) is rewritten as
\begin{equation}
R(H,\theta)=\frac{N_{\rm A} \pi^2 k_{\rm K}^2}{A_{\rm iso}g_J^2J_{\rm eff}^2\mu_B}\frac{K^{\rm qp}(H,\theta)}{\gamma_{\rm el}(H,\theta)},
\end{equation}
 where $K^{\rm qp}(H,\theta)=K_{\rm II}^{\rm iso}+K_{\rm II}^{\rm ind}(H,\theta)$ and $\gamma_{\rm el}(H,\theta)$ is determined based on the magnetic field dependence of the specific heat\cite{Shimizu2}.  $R(H)$ for $H\|[001]$ and $H\|[111]$ are presented in Fig. 5(c).   Notably, $R(H)$ exhibits a significant increase for $H \|[001]$ compared to $H\|[111]$.  }

Such an enhancement of magnetic correlations is often associated with proximity to a quantum critical point (QCP). In periodic Kondo lattices, critical behavior can occur around a characteristic magnetic field, $H_K$, where heavy quasiparticles undergo an itinerant-localized crossover.  The Kondo temperature of UBe$_{13}$ is reported as $T_K\approx6-8$ K, corresponding to a magnetic field of  $H_K=0.836\times T_K\approx 5-7$ T.\cite{Ott2}  Thus, around 6 T, the heavy quasiparticle band might be expected to lighten or vanish due to a Zeeman effect in heavy electron bands.\cite{Shimizu2,Shimizu3} This could be linked to the anomalies observed in the Hall effect, the magnetic torque, the thermoelectric power, and the  KSs.  
 
 However, this seemingly contradicts the absence of anomalies in specific heat around $T_{\rm K}$  and $H_{\rm K}$. The discrepancy with the specific heat might be reconciled in the context that UBe$_{13}$ has multiple $T_K$s associated with multiple Fermi surfaces. The heaviest Fermi surface with a low $T_{\rm K}$ disappears at low magnetic fields, leading to an enhancement of magnetic fluctuations around $H_{\rm K}$. On the other hand, the moderately heavy Fermi surface with a high $T_{\rm K}$ is thought to maintain the heavy fermion state even at 6 T, and thus would not cause significant changes in bulk probes such as specific heat and magnetization. It is considered that such multiband effects are related to the anomalous behavior of the  superconducting upper critical field, $B_{\rm c2}$, in the SC state as well as the peculiar non-Fermi liquid state in the normal state. Therefore, the hypothesis of multiple Fermi surfaces with different $T_K$s provides a compelling framework for understanding the complex magnetic field dependence observed in UBe$_{13}$. In order to test and refine this understanding, further high-field experiments would be crucial.

\begin{figure}[htbp]
     \centering
     \includegraphics[width=1\linewidth]{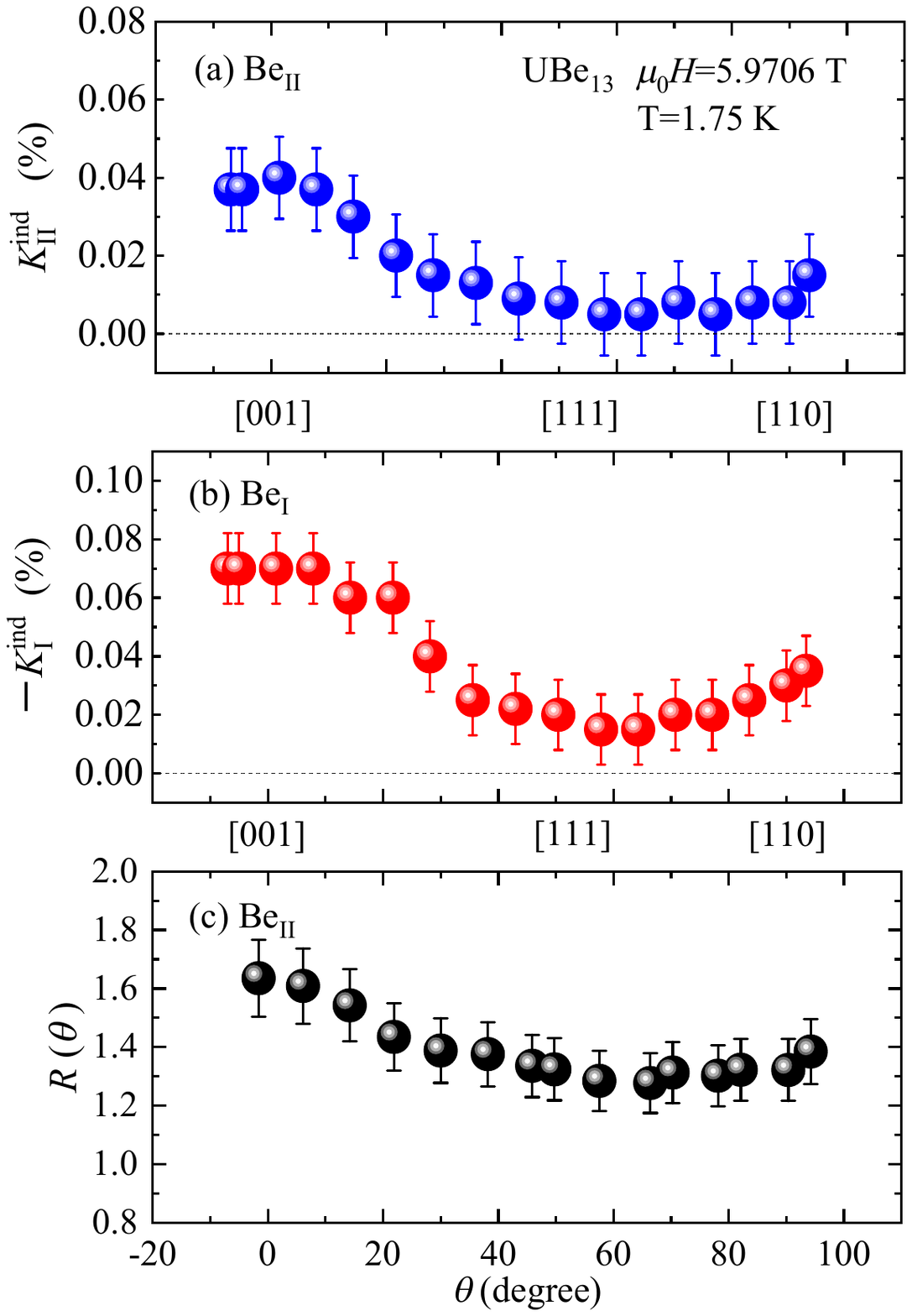}
     \caption{(Color online)
Magnetic field angle dependences of the field-induced component of the KS for (a) Be$_{\rm II}$ site and for (b) Be$_{\rm I}$ site. (c) Magnetic field angle dependence of the Wilson ratio $R(\theta)$ from $H\|[001]$ to [111].
 }
     \label{ir_spc_K}
\end{figure}

\subsection{Emergence of the field induced magnetic correlation around $H_{\rm K}$}

Finally, we address the field-orientation-dependence of $K^{\rm ind}$ at 6 T. Figure 6 illustrates the angular dependence of $K^{\rm ind}$ for the Be$_{\rm I}$ and Be$_{\rm II}$ sites at 6 T as the magnetic field is rotated in the (110) plane from the [001] to [110] directions. 

As mentioned before, at $\approx$0.5 T, $K^{\rm ind}\approx 0$ \% for both sites within the experimental error, which is consistent with  the framework of the heavy Fermi liquid model. However, at $\approx$6 T, proximate to $H_K$, a significant angular dependence emerges in  $K^{\rm ind}$. This field-orientation-dependence is characterized by a maximum for $H \| [001]$, a minimum for $H \| [111]$, with a slight recovery for $H\|[110]$.  Notably, when $H \| [111]$, the $K^{\rm ind}$ is hardly enhanced even at $\approx$6 T for both Be$_{\rm I}$ and Be$_{\rm II}$ sites within experimental error. The present data indicate an anomalous enhancement of the KS for $H \| [001]$ at $\approx$6 T. Such field-angle anisotropy is also observed in the specific heat above $\approx$2 T, where $\gamma_{\rm el}(H,\theta)$ is maximum for $H \| [001]$ and minimum for  $H \| [111]$.\cite{Shimizu2} The behavior is in qualitative agreement with the field-orientation-dependence of the field induced component of the KS. This consistency suggests an anisotropic mass enhancement, likely reflecting the Fermi surface topology.  {The field angle dependence of the Wilson ratio $R(\theta)$  is deduced from Equation (9) at  6 T and results are shown in Fig. 6(c). We found that  $R(\theta)$ is enhanced around $H \| [001]$ at 6 T, indicating enhanced magnetic correlations along this direction.}

 The underlying mechanism for the enhanced magnetic correlations along $H \| [001]$ at 6 T compared to $H\| [111]$ remains unclear. Specifically, in cubic $O_h$ symmetry, the $\langle100\rangle$ direction is fourfold rotational axis and represents the highest symmetry direction.  In such a symmetric system,  the magnetic responses along the $x, y,$ and $z$ axes are equivalent and therefore, in the absence of an external magnetic field, magnetic fluctuations are present. Even when $H \| [001] $, the symmetry within the $x-y$ plane perpendicular to the [001] axis is preserved. This means the degeneracy of the magnetic response in the $x-y$ plane directions is not lifted, and magnetic fluctuations may persist. In fact, neutron scattering experiments revealed short-range antiferromagnetic correlations with wave vector $\vec{q}=\langle 0.5, 0.5, 0\rangle$ and a longitudinal moment modulation in the normal state in zero applied field.\cite{Coad,Hiess}  Applying a magnetic field along the [001] direction does not break the symmetry of the above antiferromagnetic correlation. Therefore, the large magnetic response observed for $H \| [001]$ is not contradictory to the magnetic fluctuations from the short-range order observed by the neutron scattering experiments. Thus,  the observed maximum response along the $\langle100\rangle$ direction suggests that such an antiferromagnetic correlation arising from magnetic dipole contributions from U ions, potentially involving higher-order magnetic multipoles,  may become prominent near $H_{\rm K}\approx 5-7$ T upon application of the magnetic field. Further experimental and theoretical investigations at high magnetic fields  are necessary to elucidate the anomalous magnetic response in UBe$_{13}$.

\section{Conclusion}

In conclusion, we have reported the results of $^9$Be NMR measurements on a single crystal of the heavy fermion superconductor UBe$_{13}$.  In our 2007 study, parameters were determined that could reproduce the NMR spectra at low magnetic fields. However, they failed to reproduce the angular dependence at high magnetic fields. In this study, we performed detailed simulations considering the non-symmorphic space group of UBe$_{13}$, and determined parameters that accurately reproduce the complex NMR line profiles at high magnetic fields. In particular, the influence of classical dipole fields was found to be significant. The comparison between the KS and the classical dipolar shift provides experimental signature of itinerant-localized duality in the heavy fermion system. The magnetic field dependence of the KSs exhibits anomalies at around 6 T, suggesting that a reconstruction of a part of the multiple Fermi surfaces occurs in the high magnetic field region.  Furthermore, we observed an enhancement of the magnetic correlation around the [001] axis at $H_{\rm K}\approx 5-7$ T. This implies that magnetic fluctuations involving higher-order multipole fluctuations inherent to the cubic system play a role in UBe$_{13}$. 

\section*{Acknowledgements}


 This paper is dedicated to the memories of Prof. Kunisuke Asayama and Mr. Rintaro Matsuki. Prof. Asayama provided me with invaluable advice as I conducted NMR research. Mr. Matsuki measured the magnetic field dependence of NMR on UBe$_{13}$ from 2020 to 2022, leaving behind important results that form the core of this paper, before passing away at the young age of 26. This work was supported by JSPS KAKENHI Grant Numbers JP23K03302, JP23K03332, JP23K25829, JP23H04871 , JP15H05745, and JP15H05882.  We thank Research Facility Center for Science and Technology of Kobe University for supplying liquid helium and nitrogen.


\end{document}